\documentclass [12pt,twoside]{article}           
 \usepackage[left,running]{lineno} 
\usepackage{setspace}

\countdef\arxiv=201

\ifnum\arxiv=0 \input cpreb.tex \fi

\ifnum\arxiv=1

\usepackage{epsfig,times,lscape}
\usepackage[usenames]{color}

    \ifnum\count102=1

    \fi

    \ifnum\count102=2
\fi

\newcommand{\nc}{\newcommand}

     \ifnum\count101=1
\nc{\qI}[1]{\section{{#1}}}
\nc{\qA}[1]{\subsection{{#1}}}
\nc{\qun}[1]{\subsubsection{{#1}}}
\nc{\qa}[1]{\paragraph{{#1}}}

\def\qpar{\vskip 2mm plus 0.2mm minus 0.2mm}
\def\qL{\hfill \break}
     \fi 

      \ifnum\count101=2
 \nc{\qI}[1]{\parindent=0mm \vskip 8mm 
{\centerline{\LARGE \color{red}#1}}\vskip 3mm}
\nc{\qA}[1]{\vskip 2.5mm \noindent 
{{\bf\large\color{blue}  #1}} \vskip 1mm \parindent=0mm}
 \nc{\qun}[1]{\vskip 1mm \noindent {\sl \color{blue} #1 }\quad }

\def\qL{\hfill \break}
\def\qpar{\vskip 2mm plus 0.2mm minus 0.2mm}

      \fi

\def\qth{\vrule height 12pt depth 0pt width 0pt}
\def\qtb{\vrule height 0pt depth 5pt width 0pt}

\nc{\qfoot}[1]{\footnote{{#1}}}

      \ifnum\count101=1
\def\qbu{\hfill \par \hskip 6mm $ \bullet $ \hskip 2mm}
\def\qee#1{\hfill \par \hskip 6mm (#1) \hskip 2 mm}
      \fi
      \ifnum\count101=2
\def\qbu{\hfill \par \hskip 4mm $ \bullet $ \hskip 2mm}
\def\qee#1{\hfill \par \hskip 4mm (#1) \hskip 2 mm}
      \fi

\def\qparr{ \vskip 1.0mm plus 0.2mm minus 0.2mm \hangindent=10mm
\hangafter=1}

     \ifnum\count101=1 
 \def\qdec#1{\parindent=0mm\par {\leftskip=2cm {#1} \par}}
     \fi
     \ifnum\count101=2
  \def\qdec#1{\parindent=0mm \par {\leftskip=1cm {#1} \par}}
  
  \def\qcitb#1{\noindent \hbox to 102mm{\hfill \small #1} \vskip 1mm}
      \fi

 \def\qpages#1{\count102=0{\loop\advance\count102 by 1
 \null \vfill\eject \ifnum\count102<#1 \repeat}}

\def\qth{\vrule height 12pt depth 0pt width 0pt}
\def\qtb{\vrule height 0pt depth 5pt width 0pt}

\def\qv{\vskip 0.1mm plus 0.05mm minus 0.05mm}
\def\qhu{\hskip 0.6mm}
\def\qhv{\hskip 3mm}

\def\qhw{\hskip 1.5mm}
\def\qleg#1#2#3{\noindent {\bf \small #1\qhw}{\small #2\qhw}{\it \small #3}\qv }
\newcommand{\promille}{%
  \relax\ifmmode\promillezeichen
        \else\leavevmode\(\mathsurround=0pt\promillezeichen\)\fi}
\newcommand{\promillezeichen}{%
  \kern-.05em%
  \raise.5ex\hbox{\the\scriptfont0 0}%
  \kern-.15em/\kern-.15em%
  \lower.25ex\hbox{\the\scriptfont0 00}}

\fi

\begin{document}
\thispagestyle{empty}



\markboth{{\sl \hfill  \hfill \protect\phantom{3}}}
        {{\protect\phantom{3}\sl \hfill  \hfill}}

\color{yellow} 
\hrule height 10mm depth 10mm width 170mm 
\color{black}

 \vskip -13mm   

%
%

\centerline{\bf \large Magnitude and significance of the 
 peak of early embryonic mortality}
\vskip 7mm 
\vskip 10mm

\centerline{\normalsize
Qinghua Chen$ ^1 $,
Zengru Di$ ^2 $,
Eduardo M. Garcia-Roger$ ^3 $,
Hui Li$ ^4 $,
Peter Richmond$ ^5 $,}
\qL
\centerline{\normalsize
Bertrand M. Roehner$ ^6 $}

\vskip 5mm
\large

%
                              
\vskip 5mm
\centerline{\it \small Version of 25 February 2020}
\vskip 3mm

{\small Key-words: Embryogenesis, death rate, conception, infant death, 
senescence} 

\vskip 3mm

{\normalsize
1: School of Systems Science, Beijing Normal University, China.\qL
Email: qinghuachen@bnu.edu.cn\qL
2: School of Systems Science, Beijing Normal University, China.\qL
Email: zdi@bnu.edu.cn\qL
3: Institut Cavanilles de Biodiversitat I Biologia Evolutiva,
University of Val\`encia, Spain.\qL
Email: eduardo.garcia@uv.es\qL
4: School of Systems Science, Beijing Normal University, China.\qL
Email: huili@bnu.edu.cn\qL
5: School of Physics, Trinity College Dublin, Ireland.\qL
Email: peter\_richmond@ymail.com \qL
6: Institute for Theoretical and High Energy Physics (LPTHE),
Pierre and Marie Curie campus, Sorbonne University,
National Center for Scientific  Research (CNRS),
Paris, France. \qL
Email: roehner@lpthe.jussieu.fr\qL
}
\vfill\eject

%

\large

{\bf Abstract}\qL
Biologically,
for any organism life does not start at birth but at fertilization 
of the embryo.
Embryonic development is of great importance because 
it determines congenital anomalies and influences their
severity.
Whereas there is detailed qualitative knowledge of the successive steps 
of embryonic development, little
is known about their probabilities of success or failure. 
Embryonic mortality
as a function of post fertilization time provides a simple (albeit crude)
way to identify major defects. We find that, in line with 
the few other species for which data are available,
the embryonic mortality of zebrafish has a prominent peak
shortly after fertilization. This is called the early embryonic
mortality (EEM) effect.
Although a number of immediate causes of death
(e.g. infection, excess of carbon dioxide or of lactic acid,
chromosomal defects) can be cited, the
common underlying factor remains unknown. 

After reviewing embryonic mortality data
available for chicken and a few other farm animals,
we explain that zebrafish are particularly suited for
such a study because embryogenesis can be followed 
from its very beginning and can be observed easily thanks
to transparent egg shells. 

We report the following findings.
(i) The mortality peak occurs in the first 15\% of the
75-80 hours of embryogenesis and it is about 50 times higher than
the low plateau which follows. (ii) The shape of the age-specific
death rate is largely independent of the death level.

Presently, little is known about the nature of embryonic  defects.
However, by reviewing two special cases we show
that even small initial defects, e.g. spatial cellular
asymmetries or irregularities in the timing of development,
carry with them lethal effects in  later stages
of embryogenesis.

\vfill\eject

\count101=0  \ifnum\count101=1

\vfill\eject

\fi

\large

\qI{Introduction}

\qA{From defects in embryonic cell organization to lethality}

The purpose of statistical physics is to derive the behavior
of macroscopic systems from the collective properties
of its microscopic components. The present investigation 
gives a striking illustration for it
shows a massive mortality surge in early embryonic development
which comes along with slight disturbances at cellular level.
More specifically, investigations by two teams of 
researchers (Rideout et al. 2004, Cruz et al. 2012)
have shown that anomalies in initial cell organization
are early signals
of a compromised (and eventually lethal) development.
It is interesting to note that in these two cases
(to be explained more fully later on) the defects are not
due to individual cells but rather to 
defects in their interaction. 
Note that the defect considered in Rideout et al.
is a spatial asymmetry that is fairly similar to
positional defects in a crystallized solid.
\qpar

In line with some of our previous papers (Berrut et al. 2016,
Richmond et al. 2019a,b, Bois et al. 2019b) 
this study could have been
entitled ``The physics of embryogenesis'' 
because all along we keep in mind possible parallels
with the working and failures of physical (i.e. non living) systems;
such parallels serve as guides for the exploration
and suggest possible conjectures.
We do not try to give a detailed description
for a single species but focus on common features
across species for it appears that
similar peaks occur in many species;
in short, we study embryonic mortality
as a global effect.
\qpar

So far, limited mortality data are available. Those recorded 
in our experiment for
zebrafish embryos may be the most accurate
ever produced. There are two reasons for that.
The experiment was especially designed for the observation
of early embryonic mortality (whereas previous experiments 
with farm animals had mostly an economic goal)
and it involved a sample of
over 1,500 embryos. It is difficult to
have samples of this size for bigger animals like 
cows or pigs.
\qpar

The early embryonic spike is not the only mortality peak 
to occur in the development of an organism. In order to
put it in a broader perspective, in the following subsections
we briefly describe the other peaks.

\qA{From fertilization to death there are three phases}

In previous papers a parallel was introduced between living and
technical systems (Berrut et al. 2016,
Bois et al. 2019a,b). From this perspective 
three phases can be distinguished in human development.
%
\begin{figure}[htb]
\centerline{\psfig{width=18cm,figure=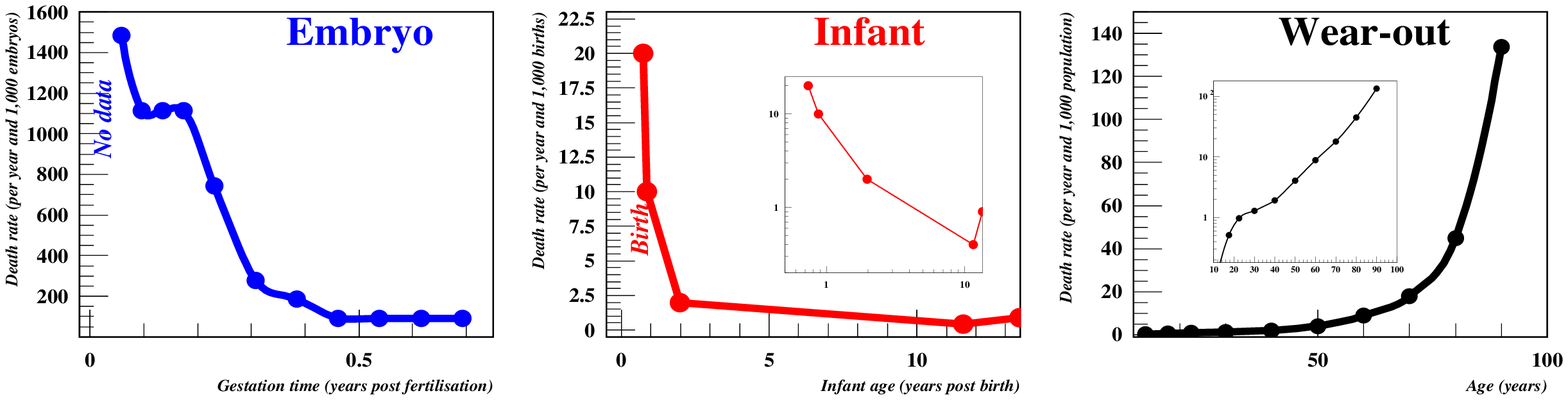}}
\qleg{Fig.\qhu 1\qhv Evolution of the death rate in the three phases 
of human development.}
{In each phase there is a huge peak. Whereas its existence is easy to understand in
phase 2 and 3, the reason of the early embryonic mortality (EEM) peak
remains a mystery. In phase 2 the decrease follows
a hyperbolic power law whereas in phase 3 the increase 
is exponential as described by Gompertz's law (1825); 
in phase 1 the exact shape of the curve is not yet known
due to a lack of accuracy in empirical evidence, in particular 
in the first few weeks following fertilization. The expression
``wear-out phase'' is borrowed from reliability engineering; it is justified
by the fact that for many functions the best performance occurs between 
the age of 20 and 30 and then deteriorates steadily. As examples one can
mention the ability to perceive sounds of high frequency or the density of bones.
In most activities of common life this wear-out remains unnoticed and becomes
a hindrance only in old age. 
Note that the maximum death rate
of the senescence curve depends upon the the oldest age group selected. A similar
observation applies to the infant curve; in this case inclusion of
age groups close to birth would lift up the maximum of the death rate.
The data for phase 1 are for the island of 
Kauai in Hawaii in the late 1950s. The data for phase 2 are for the United States in 1960.
The data for phase 3 are for the United States in 2016.
The insets of panel 2 and 3 are (log,log) and (lin,log) respectively.}
{Sources: 
Embryo phase: French et al. 1962, L\'eridon 1973,1977, Jarvis 2017.
Infant phase: The data from zero to 12 years are presented in two different tables
(a) Under one year: Grove et al. 1968, p.210-211, 
(b) Over one year: Grove et al. 1968, p.318.
Wear-out phase: WONDER database of the ``Centers for Disease Control'' 
(CDC), Compressed mortality, 1999-2016.}
\end{figure}
%
\qee{1} Embryogenesis from fertilization of the embryo to birth.
\qee{2} From birth to 10 year old there is the infant phase%
\qfoot{The medical definition of infant mortality is mortality of newborns
under one year. This definition is mostly motivated by statistical
convenience (a more detailed discussion can be found in Berrut et al. 2016,
appendix A). From a biological perspective it is more
logical to include into the infant phase the whole age interval during
which the death rate decreases which is 0-10 years.
Note also that
the "beginning" of the death rate curve has no
real signification for the following reason.
The first point is for age less than
one year. However, if we would take as first point
the mortality rate for age less than one
month, or less than one day, or less than one hour,
the rate would be at least 100 times higher. 
More details about this
hyperbolic behavior can be found in one of our earlier
papers (Berrut et al. 2016). }%
, 
also known as a wear-in phase.
\qee{3} After the age of 10 year and until senescence and death comes 
a phase during which the death rate increases exponentially
in accordance with the well-known Gompertz's law 
(Gompertz 1825, Richmond et al. 2016).
In the language of reliability engineering such
a steady increase of the failure rate is known as
a wear-out process.
\qpar

Human age-specific curves for the three phases are given in Fig.1.
For the sake of brevity in what follows the three phases will be
referred 
to as phases 1, 2 and 3. 
Whereas there are accurate data for phases 2 and 3,
embryonic death rates, and especially the early embryonic rates 
are subject to a number of uncertainties as
emphasized in the review paper by Gavin Jarvis (2017).
The main reason is the difficulty of detection
of fertilization in the first two weeks (e.g. early
hormonal dosage is subject to several external conditions).
Similarly,
early fetal deaths go unrecorded. For instance, in the United States
only fetal deaths at 20 or more weeks of gestation are recorded
by states in the ``National Vital Statistics System''.
This situation suggests two comments
\qbu
Despite the statistical uncertainties of human embryogenesis, 
because of its sheer magnitude (multiplication by 7 with
respect to the base line), it is clear  that there
is a major mortality surge in the early phase of embryogenesis.
This is called the early embryonic mortality (EEM) effect.
\qbu This effect can be studied in much better conditions in fish than
in mammals or birds because in the former fertilization occurs outside
the body of the female and can be followed from the very beginning.
Zebrafish eggs have the additional 
advantage that their shells are transparent, a feature
that is not common to all fish, e.g. the egg shells of the
killifish, another model fish used in laboratories,
are {\it not} transparent.

\qA{Death rate surges in phases 2 and 3}

In the embryonic and infant phase the mortality surge
is at the start whereas in the senescence
phase it is at the end. Coming back to humans
cases 2 and 3 are fairly easy to understand.
\qbu  In phase 2 the newborns
face many challenges such as breathing, regulating body temperature,
being able to absorb and digest the mother's milk. 
An inability to perform 
any one of these crucial functions means that the newborn 
may not live very long.
Defective lungs which were acceptable during pregnancy due to the
oxygen supply provided by the mother may not be able to support the life
of the newborn after birth. 
Thus, birth appears to
act as a filter resulting in the loss of
newborns afflicted by
life threatening defects. In spite of substantial
progress in neonatal assistance, present-day data still
show a marked post-natal mortality surge. 
In fact, mortality during the first few
days after birth did not change much over past decades
This is because major cardiac or neurological
defects can hardly be corrected. 

Note that, in contrast to Gompertz's law, the post-natal fall
is not an exponential but a power law 
(more detailed graphs can be found in Berrut et al. 2016).
\qbu The mortality surge seen in
phase 3 is not a filtering effect. In engineering language 
it is known as a wear-out effect.
According to Gompertz's law the death rate basically
doubles every 10 years of age. 
With a logarithmic vertical scale the graph is a straight line
(as shown in the inset of Fig.1, panel 3);
with a linear vertical scale it appears
in the form of a sharp increase. Gompertz's law implies that
the size of a cohort decreases as an exponential whose
exponent (i.e. the death rate) increases itself exponentially
with age.
Although it would be interesting to investigate the reasons
of such a very abrupt decline, this study would bring
us too far away from our main topic
and it will rather be postponed to an upcoming paper.
\qpar

In the following sections the paper proceeds as follows.
\qbu In addition to the human data already discussed in the introduction,
we present embryonic mortality data for the few other cases that have
been studied.
\qbu We explain the design of our zebrafish experiment and we discuss
its results.
\qbu In the conclusion we extend the 3-phase pattern given in the
introduction.  We explain also how embryonic mortality can 
be used to probe the temperature dependence of the successive
biochemical processes of embryonic development. 
This methodology is the 
parallel of the injection methodology for the exploration of the 
DNA structure.
\qbu Finally, appendix A discusses some basic practical aspects
of the experiment (e.g. the oxygen supply issue).

\qI{Insights into the embryonic mortality surge}

The literature gives us almost no real clues for understanding
the EEM effect. As
our only solid starting points one can recall the following facts.
(i) Far from being specific to humans, such early spikes
seem to exist for all species for which data are available
(see below).
(ii) Most often autopsies are unable to identify the 
immediate causes of fetal deaths. Even in those cases
where specific causes are found, they appear
fairly disparate (e.g. bacterial or viral infections, lack of oxygen,
accumulation of ammonia or lactic acid, chromosomal anomalies)
and do not reveal the common effect which may lie behind.

Even if one accepts the fairly natural
idea (borrowed from the neonatal phase) 
that the EEM is a filtering process (which
in reliability engineering is called a wear-in process)
this gives little insight. 
Contrary to birth defects, one cannot point
to failure of specific organs for in the early embryonic phase
(i.e. the cleavage phase) there are only undifferentiated
cells.  That is why we think it may be enlightening to describe
two cases in which defects in early cell organization
carry lethal effects.

\qA{Effect of spatial asymmetry in the cleavage phase}

Fig.2 shows position anomalies which appear at the 8-cell
level that is to say, assuming a 20mn division time,
some $ 60 $ mn after fertilization. This asymmetry
comes together with unequal cell sizes. 
%
\begin{figure}[htb]
\centerline{\psfig{width=15cm,figure=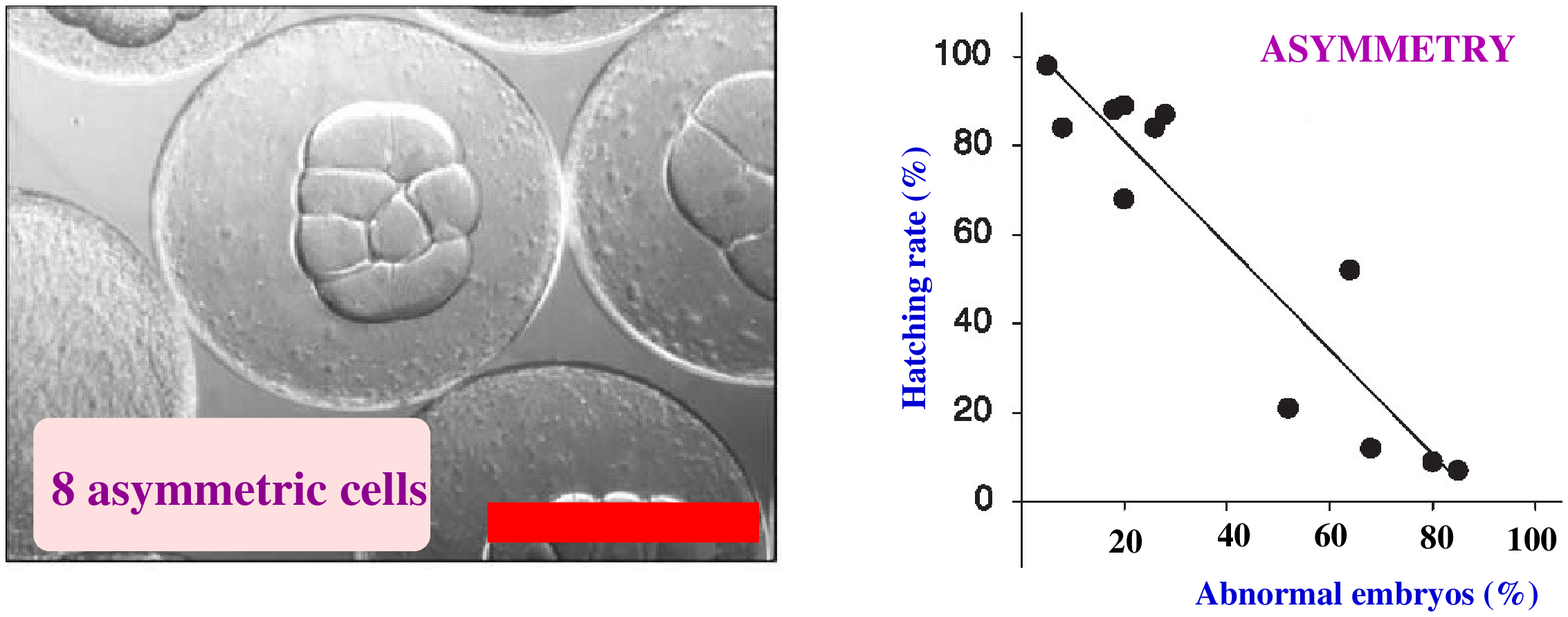}}
\qleg{Fig.\qhu 2\qhv A position asymmetry in haddock embryos
carries a lethal outcome.}
{Other kinds of cellular anomalies at the same 
development stage are described in the same paper.
It is remarkable that cellular outcrops where one or two cells
protrude from the same cluster of cells do not have
any negative effect in terms of hatching. The red segment
corresponds to 1mm.}
{Source: Adapted from Rideout et al. (2004).}
\end{figure}
%

The graph shows 
that almost all embryos affected by such a defect will die
before hatching time. It would of course be of great interest
to know the average time lag between the apparition
of this defect and the death of the embryo. The authors
of Rideout et al. did not study this aspect because
the paper was written for fisheries and
for obvious economic reasons it was the hatching rate 
which was their main concern. In a subsequent paper
we intend to study this aspect.

After this example of spatial asymmetry we give
an example of temporal asymmetry.

\qA{How early timing irregularities affect the viability
of the embryos}

If all cells replicate in a synchronous way 
development leads successively to 1, 2, 4, 8, 16, 32, 
64, $ \ldots $ cells%
\qfoot{In fact the divisions of the zygote (i.e. the maternal
ovum after its fertilization) are not the first divisions
of the maternal cell.
The oocyte has already the ability to divide. 
Under the influence of reproductive hormones a primary oocyte
completes a meiotic cell division at the end of which it splits 
into two separate cells: a small one which is fairly useless
and a large secondary oocyte. Needless to say, this
division can also be faulty.}%
.
However, observation shows deviations from synchronicity,
a fact that is not surprising actually.
As they have the same DNA the daughter cells are clones of the
initial zygote. 
However, it is known that even for clones there is a dispersion
in division time (Walden et al. 2016, Bois et al. 2019b). 
This means that at some moments in
this replication process there may be 3, 5, 6 or 7 cells 
(instead of 2, 4 and 8).

Will such a dispersion
affect the subsequent development of the embryo? 
An observation based on
834 human embryos shows that the answer is yes (Cruz et al. 2012). 
It 
appears that for the 552 embryos which had a successful development%
\qfoot{In this {in vitro} fertilization ``successful'' means 
that the embryo reached a stage (called
blastocyst stage) which is just prior to its implantation 
in the uterus.}
the time spent
in a 3-cell stage was on average 0.6 hour, whereas for the 282 unsuccessful 
developments
the time spent in a 3-cell stage was one hour. All other parameters describing
the timing of the divisions were the same within a $ \pm 10\% $ margin. 
In other words, this observation suggests that a lack of synchronicity 
in cell divisions 
had a disturbing influence on embryonic development.
\qpar

Here the effect is less massive than in the observation
described in the previous subsection. Clearly, it would
be quite important to repeat this experiment with non-human embryos,
for instance zebra embryos. Although early embryonic
developments in humans and fish are fairly different,
one would expect a similar effect for the simple 
reason that an embryo with 5 cells has necessarily a spatial
asymmetry. Even if only temporary
it should affect further development.

\qI{A testable conjecture for the embryonic mortality peak}

Our idea of a parallel between embryogenesis and 
a manufacturing process may appear somewhat speculative.
However, the nice side of it is that it can be tested.
Here we propose a testable conjecture.
It should be noted that it is only one of several
possible conjectures that can be derived from such a 
parallel.
\qpar

In reliability engineering it is a common belief
that, assuming identical technical equipment, 
manufacturing defects increase with the speed
of  the production process. The reason is simple.
Control procedures take time (even if they are
automatic). In successive biochemical reactions
control mechanisms may be set up after each step.
Such controls improve overall reliability but
at the cost of slowing down the process.
\qpar

It turns out that across species there are great differences
in early division times:  0.5 hour for {\it C. elegans}
(a common laboratory model organism), 
0.5 hour also for zebrafish embryos,
10 hours for mice, 20 hours for cows, 24 hours for human
embryos. According to the previous argument,
the hatching rates of different yet "similar" species
should be in proportion of their initial division times.
\qL
Clearly, the main difficulty is how to define 
the notion of ``similar'' species.
For the sake of simplicity we retain the number of cells 
at hatching. This may be a fairly rough criterion
but at least it has the advantage of being well defined;
e.g. {\it C. elegans} has about 1,000 cells whereas zebrafish
have some 25,000 cells. Thus, this argument would
lead us to predict a lower hatching rate for
zebra fish than for {\it C. elegans} larvae.
More generally, the conjecture can be stated as follows.

\qdec{\it {\color{blue} Mortality conjecture}. \qL
For two species having approximately the same number of cells
at ``birth'', early embryonic mortality
should decrease when the initial division time increases.
}

Once more data become available, it will 
be possible to test the conjecture. If not
confirmed, it may be that
our criterion based on the number of cells
is too crude. For instance, cell diversity 
may also play a key-role.

\qI{Embryonic mortality in farm animals}

For humans (and more generally for mammals)
a distinction is made between embryonic
development and fetal development. 
Because we will mainly be
concerned with zebrafish for which there is 
no real need for such a distinction,
we will homogenize our terminology and use
``embryonic'' for the whole process from fertilization to birth.
\qpar

The two main sources from which data 
can be expected are studies of farm animals
(e.g. poultry or cattle) or of
organisms used in laboratories (e.g. {\it C. elegans} or mice).

\qA{Embryonic death in chickens}

The duration of embryonic development from fertilization to 
hatching is about 21 days.
Fig. 3a gives the graph of embryonic mortality based on data given in
Pe\~nuela et al. (2018).\qL
In this study the number of deaths in each age interval was obtained in two ways.
(i) Candling (that is to say screening the egg with the help of a light source).
(ii) After hatching time the eggs which did not hatch were opened and the day
of death was established.
At that point the distinction between non-fertilized and
fertilized eggs could be made easily: any egg which
experienced embryonic development, no matter how small,
could be considered fertilized.
\qpar
%
\begin{figure}[htb]
\centerline{\psfig{width=8cm,figure=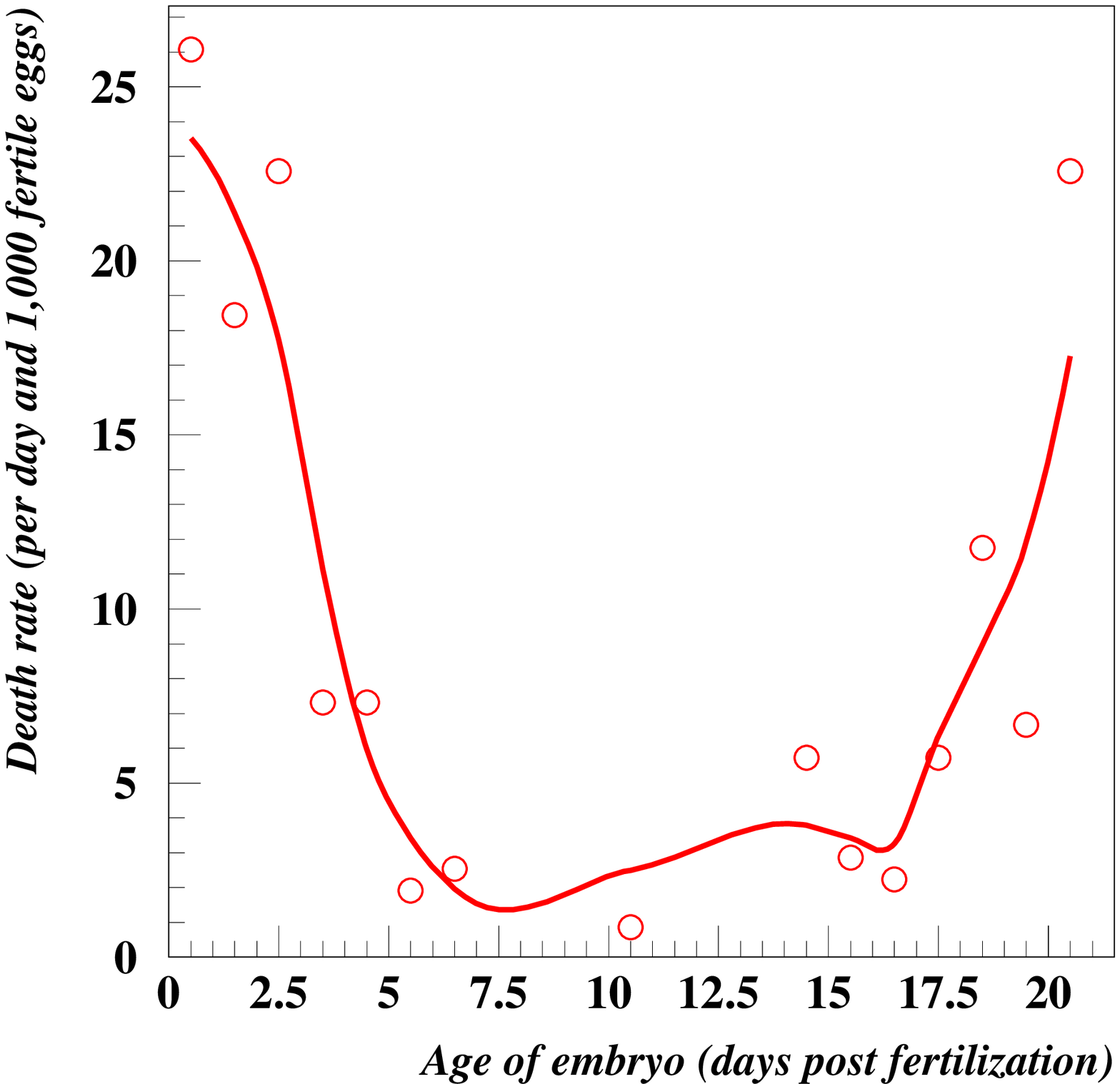}}
\qleg{Fig.\qhu 3a\qhv Embryonic death rate in chickens.}
{The definition of the death rate which is used here is
the same as for infant mortality that is to say
number of deaths in a given age interval divided by initial
number of embryos.
In this study a total of 3,240 eggs were examined and there were
a total of 471 embryonic deaths which gives a hatching rate of 82\%.
Of these deaths, 57\% occurred during the
first week. Older age of the female increased death rates but did not change
the shape of the curve.
The reasons which explain the occurrences of a maximum just prior to hatching
are explained in the text. Malposition of the embryo inside the egg
was a major cause of pre-hatching death.}
{Source: Pe\~nuela et al. (2018, p.6505)}
\end{figure}
%
Apart from the early peak there is another maximum just before hatching.
It has two explanations.
\qbu It has been observed that one half of all
chick embryos which die between day 18 and 20 were in
an abnormal position which did not give them access to the air cell 
which is on the blunt tip of the egg (Hutt 1929). 
\qbu The lungs of chicks start to work
shortly before hatching; previously the oxygen
absorbed into the egg through the shell
was transferred to the embryo through a network of thin
blood vessels constituting a kind of rudimentary lung 
called the allantois. 
If
the amount of oxygen delivered is insufficient, breaking
the shell becomes impossible.
\qpar

Because for chicks (and more generally for all avian species
whose eggs have a fairly hard shell) breaking the shell is
a very challenging task, one is not surprised to see a high
pre-hatching mortality peak. One would not expect a similar
peak in zebrafish embryos which have a fairly soft egg membrane
and this is indeed confirmed by
observation. 
However, the same post hatching (or rather post-yolk)
filtering process is expected
for fish and birds likewise.

Similar results were reported for other avian species such as:
pigeons, 
doves, ducks, grouse, pheasants, quails (Romanoff 1949)
or turkeys (Fairchild et al. 2002, p. 262, Bois et al. 2019b).

\qA{Embryonic death in cattle}

The widespread use of artificial insemination allows accurate 
determination of the
time of fertilization. Fig.3b shows that the mortality peak occurs 
shortly after fertilization. 

%
\begin{figure}[htb]
\centerline{\psfig{width=8cm,figure=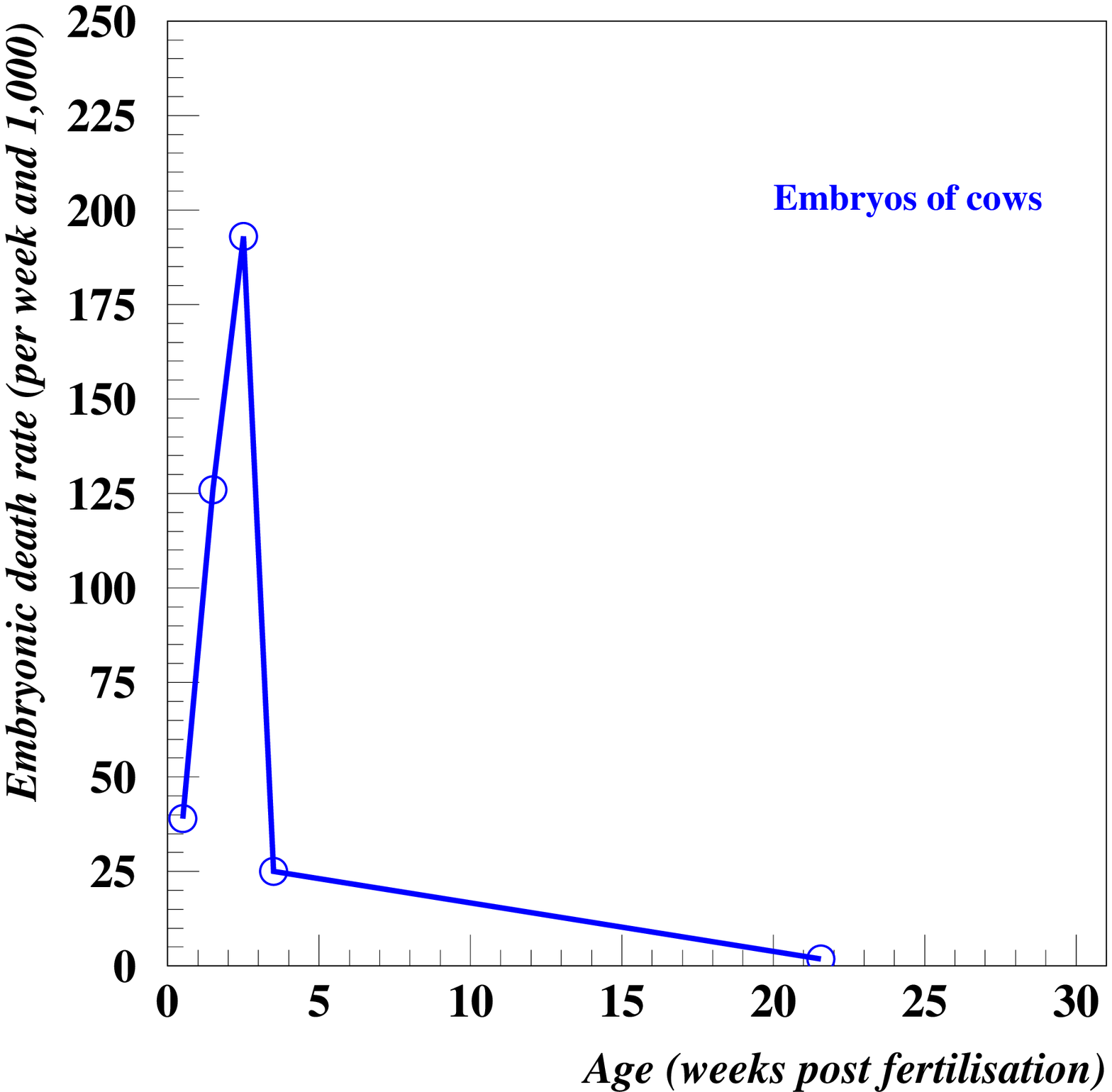}}
\qleg{Fig.\qhu 3b\qhv Embryonic death rate in cows.}
{The graph gives death rates per week of age and 1,000 successful
inseminations.
The study involved 63 cows and there were 35 embryonic deaths
which gives a calving rate of 55\%. The maximum occurred in the third week,
i.e. 15 to 21 days after fertilization.
The whole pregnancy lasts about 280 days.}
{Sources: Sreenan et al. (1986), First et al. (1988).}
\end{figure}
%
\qpar

In contrast with Fig.3a, there is no late maximum. This cannot
be seen on Fig.3b because the graph does not extend to
the end of gestation which occurs around week 40.
However, in a paper by Wathes et al. (2016) 
one learns that
from insemination to week 8 the pregnancy losses are about 50\%
of initial pregnancies
whereas the losses between week 8 and the end of pregnancy
are only 5\%.
\qpar

Predominance of early embryonic mortality is also observed in other
farm animals.  
However in this case the results are less detailed than for cows
as they give only the percentages of deaths which occur in the first third
of the pregnancy. A constant death rate throughout embryo development
would mean a share of 33\% in this first third; instead the real
percentages are as follows
(First et al. 1988):\qL
 \centerline{mare: 87\% ,\ sow: 75\%, \ ewe: 60\%} \qL
Moreover, the percentages $ p $ of total embryonic deaths as a 
fraction of successful inseminations were as follows:\qL
\centerline{mare: $ p=40\% $,\ sow: $ p=40\% $, \ ewe: $ p=25\% $}

\qA{Model organisms used in laboratories}

Regrettably, almost no mortality data as a function of embryonic age could
be found in the literature. The only available case concerns the 
European perch (Alix 2016).
It reveals  a huge initial peak (see Fig. 6b).
\qpar

It is this paucity of data which led us to explore the embryonic mortality
of zebrafish.

\qI{Exploration of the mortality of zebrafish embryos}

For a number of reasons
zebrafish offer ideal conditions for the
exploration of embryonic mortality.
\qee{1} As for most fish the eggs of the zebrafish are fertilized outside
of the body of the female. Therefore, in contrast to many other organisms
(e.g. {\it C. elegans}, rotifers, birds, mammals) the embryo 
can be observed from the very moment of fertilization to hatching.
\qee{2} As the eggs are transparent one can see easily what is going on inside.
\qee{3} The embryonic phase lasts about 3-4 days (depending on temperature)
which is a convenient duration. Shorter, it may not give enough
time to carry out the required observations. 
Longer, it would slow down the whole experiment 
without any additional benefit.
\qee{4} After death of the embryo the eggs become opaque
and therefore can be easily identified
\qee{5} Finally, as a model organism actively studied
in genetic research, zebrafish are raised in many university laboratories.
\qpar

However, we will see that despite such favorable conditions, the exploration
of embryonic mortality raises several questions.

\qA{The question of the black eggs}

In experiments involving zebrafish embryos
there are usually two preliminary
steps. Firstly, after breeding, the eggs are collected and then,
usually some
24 hours later, the eggs which have become opaque%
\qfoot{Under a stereomicroscope with light coming from below
such opaque eggs appear black, whereas with light coming from above
they appear white (see Fig.3a).
Here, we will call them ``black'' because 
this is the color which appears on the picture.}
are removed because they are dead (see Fig.4a and panel 4 of Fig.4c).
This could seem to be
a simple operation but in fact it raises some questions.
\qbu At 24 hours post fertilization (hereafter, hpf)
there will be three kinds of black eggs: (i) those
which were not fertilized and became black as a result,
(ii) those which were fertilized but were not able to develop,
(iii) those which were fertilized, developed for a few
hours but died at some point before becoming black too.
Most often it is assumed that all black eggs were not fertilized.
We will explain in a short moment how to separate (i)+(ii) from (iii).
\qbu A second question is 
what is the  time-delay between the moment when a fertilized egg
dies and the moment when it turns black?

\qA{Early separation of fertilized from unfertilized eggs}

In order to answer the first question, one should separate the fertilized
from the unfertilized embryos and do that as early as possible.
A procedure is suggested by the comparison of the
pictures shown in Fig.4b (panel 3) and 4c (panel 3). 

%
\begin{figure}[htb]
\centerline{\psfig{width=4cm,figure=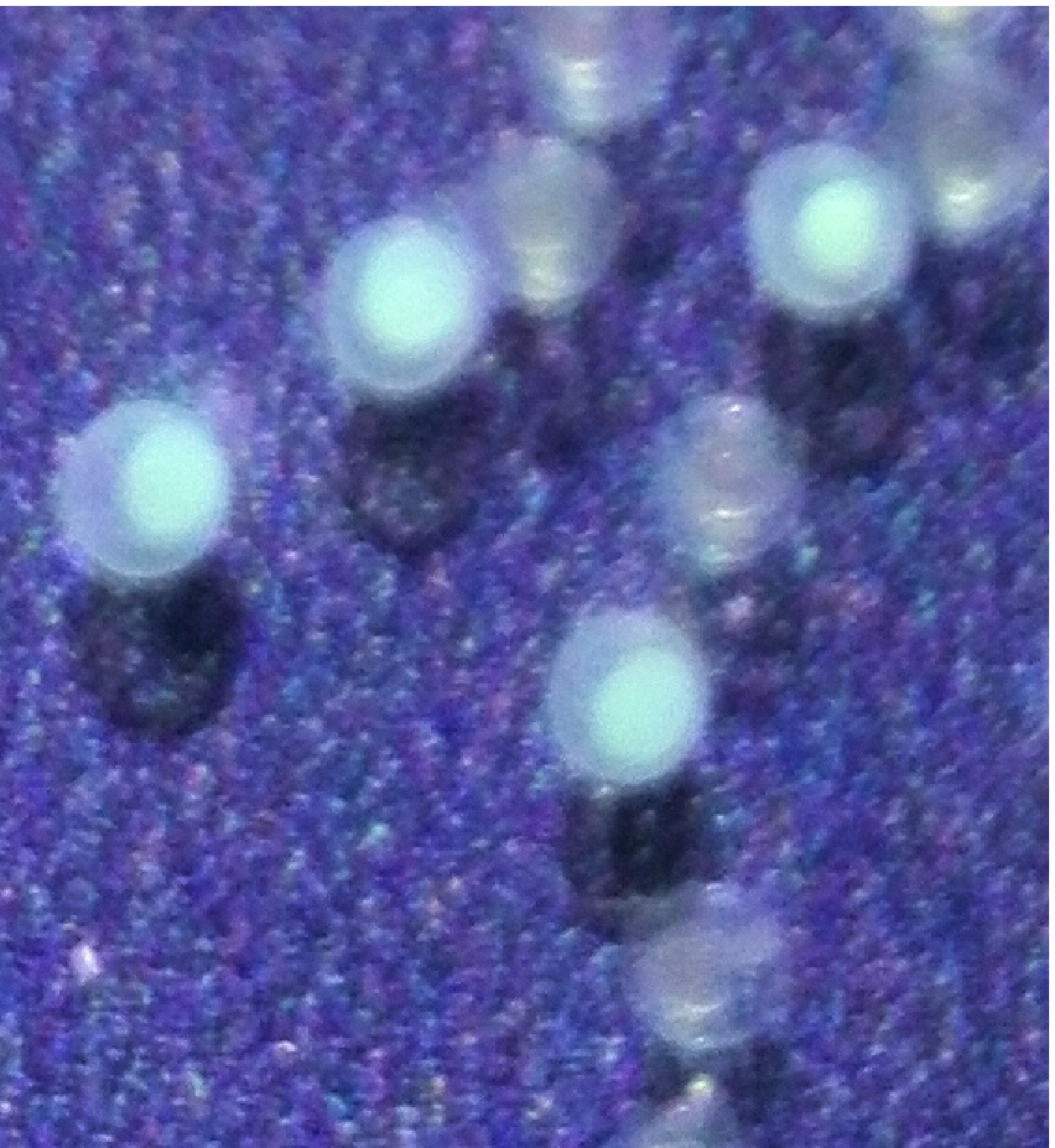}}
\qleg{Fig.\qhu 4a\qhv Live eggs (transparent)
 and dead eggs (opaque) with light from above.}
{With the light from above the opaque cells appear white.
On the contrary with light from below the sample, they appear black (as
in the last picture of Fig.3c). }
{Source: Picture taken at the ``Biology Institute of Paris-Seine''.}
\end{figure}
%

%
\begin{figure}[htb]
\centerline{\psfig{width=8cm,figure=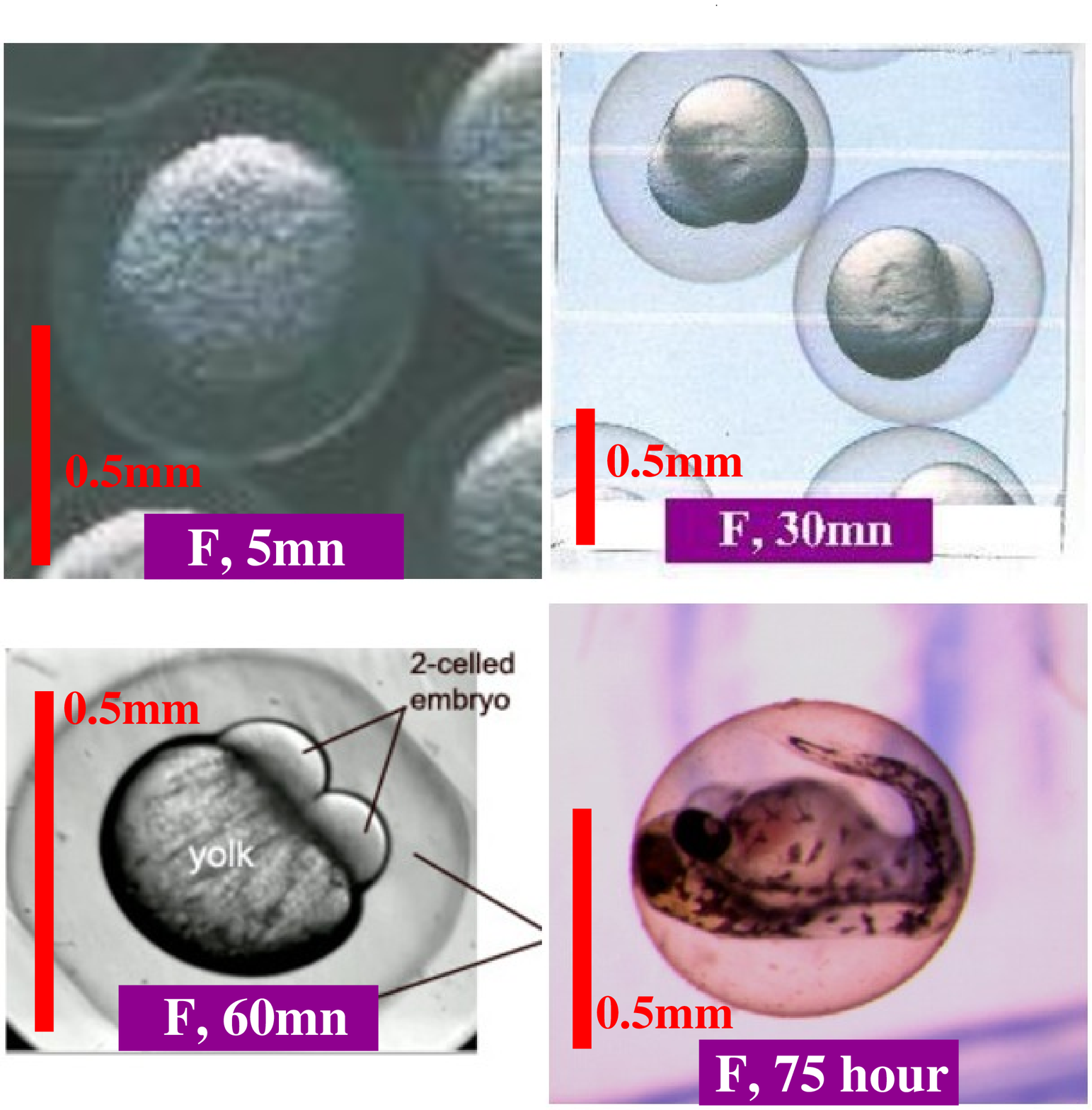}}
\qleg{Fig.\qhu 4b\qhv Successive steps in the development of fertilized embryos.}
{A critical step which occurs about one hour after fertilization
is the apparition 
of a second hump (picture 3) which results from the first
division of the initial embryo cell. The last picture shows the embryo
shortly before hatching.}
{Source: Pictures (except for panel 3)
taken at the ``Biology Institute of Paris-Seine''.}
\end{figure}
%
%
\begin{figure}[htb]
\centerline{\psfig{width=8cm,figure=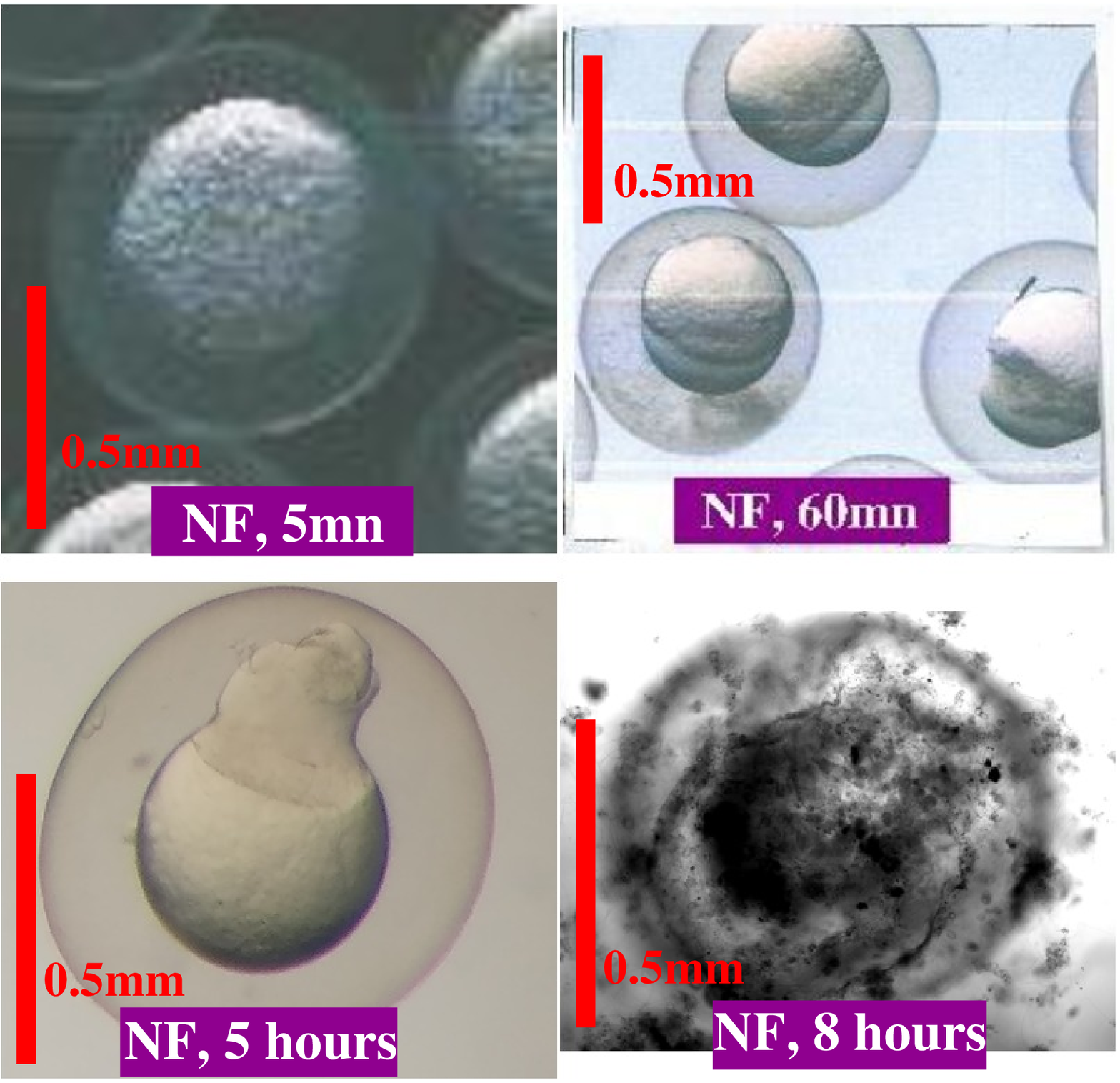}}
\qleg{Fig.\qhu 4c\qhv Successive steps in the development of
  unfertilized embryos.}
{The unfertilized eggs can be obtained directly from a female by a gentle
massage on its belly. After that the process starts when the
eggs are released in water.
In fertilized as well as unfertilized eggs
one sees the apparition of a first hump. 
Then, whereas the fertilized eggs develop a second hump
as shown in the previous figure,
in unfertilized eggs
the single hump continues to swell while at the same time taking 
fairly weird shapes 
as shown in the third picture taken some 5 hours after
release in water. 
In a process which starts some 8 hours after release  all
non fertilized eggs will turn black as shown in the
last picture.}
{Source: Pictures taken at the Institut de Biologie de 
Paris-Seine by Alex Bois 
and at ``Beijing Normal University'' by Yi Zhang.}
\end{figure}
%

The fertilized eggs can be easily distinguished from the non-fertilized
after they have developed a second hump%
\qfoot{Note that at this point it cannot be strictly excluded that 
some fertilized eggs are nevertheless too defective to develop a second hump.} 
which occurs about one hour after
the eggs were produced by the female; see Fig.4b.
However in the following hours the number of humps will
increase to 4, 8, 16, 32 and so on. At each division the new cells become smaller
because after division they are not given enough time for further growth. When there
are 32 cells (or more) they form a large hump in which individual
cells can no longer be seen
and which is not very different from the single humps of unfertilized
eggs (see Fig.5, panel 2).
In other words, the separation of unfertilized eggs should best be done between
one and two hours post fertilization. As the eggs must be examined one by one
a single operator can hardly treat more than 200 eggs per hour. For larger
samples%
\qfoot{Say, over 400; in our experiment the largest cohort numbered 770.}
the solution is to increase the number of operators or/and produce
the eggs in successive
batches by removing the separations in the breeding boxes with time lags
of about one hour.
We have been using two operators and one-hour
time lags between successive batches.
\qpar

One way to test whether the separation was done satisfactorily is to collect the
discarded eggs and to see if all of them become black after 24 hours.
In the experiment presented in the next section
our policy was to accept an egg as fertilized only if 2 or 4 humps could 
be identified clearly. Thus, when an egg was oriented in such a way
that its humps were hidden (under it when using a stereo-microscope or
or above it when using an inverted microscope) 
it was rejected even though it may have been
a fertilized egg. Such a drastic selection is necessary because it is
crucial not to include any unfertilized egg as this would
artificially inflate initial death rates. 
The fact that some 95\% of our discarded eggs
became black after 24 hours shows that the procedure worked fairly well.

\qA{Identification of dead embryos}

In following the development of the fertilized eggs the most
impressive observation 
is that 
some 6 hpf the number of new black eggs 
starts to increase, reaching a maximum before dwindling to zero around
24 hours post
fertilization. Note that there may be a time lag between the moment when an
embryo dies and the moment when the egg becomes black. 
Our observations suggest that this time-lag increases in the course
of the eggs's development. In the appendix we mention special cases for which the 
time-lag was of the order of two days.
\qpar

With respect to the identification of deaths 
there are broadly speaking 3 phases:
\qee{1} The cleavage phase which lasts about 2 hours and
during which one can count the number of cells. 
We did not see any stoppage in development in this phase.
\qee{2} After that the embryo assumes a shape on which changes are not 
easy to identify%
\qfoot{Of course, if one follows a single egg in the course of time even small
changes may be visible but in our experiment we follow a large number
of eggs and we return to the same egg only
every 5-6 hours; as in such successive observations the orientation of
the embryo changes, small differences may not be clearly visible.}%
. 
This is illustrated in Fig. 5.
%
\begin{figure}[htb]
\centerline{\psfig{width=10cm,figure=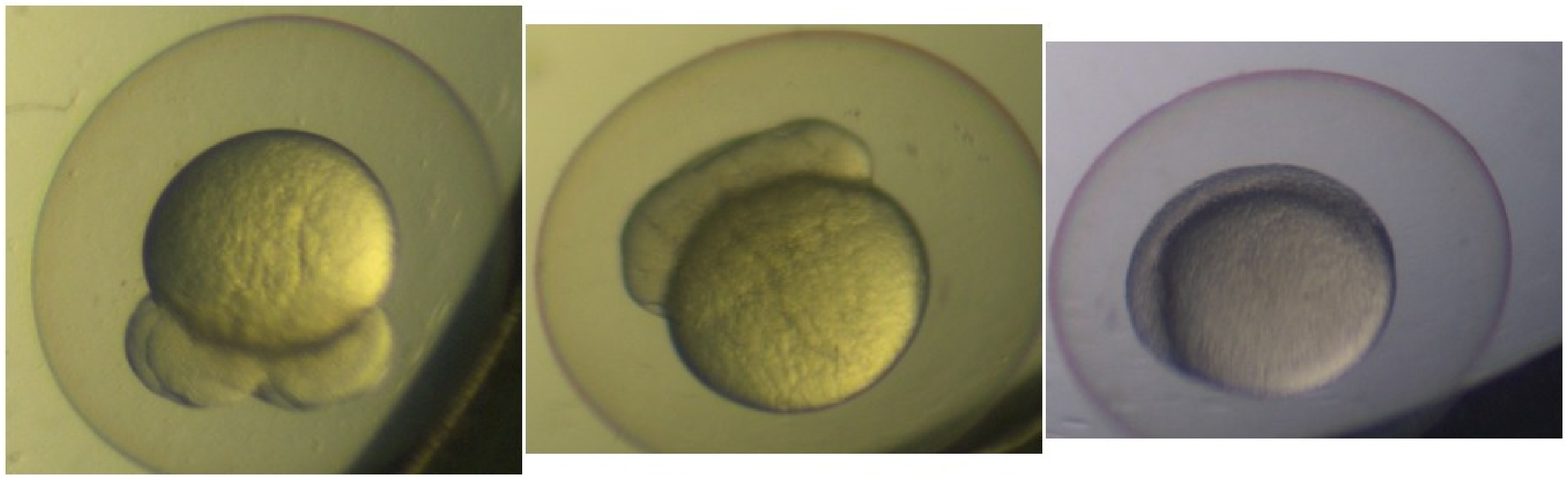}}
\qleg{Fig.\qhu 5\qhv Transition from multi-hump to thin layer.}
{The multi-hump structure results from rapid cellular division. In the first picture
there are only 4 cells whereas in the second the cells are so small and their number
so large  that one can no longer distinguish them and count them. 
In the third picture the cells form a thin
layer around the yolk ball. It is in this stage (from 4 hpf to 10 hpf)
that the evolution of the embryo is the most difficult to follow.}
{Source: Pictures taken at the ``Department of Life Science'' of ``Peking University''.}
\end{figure}
%
At this stage the most frequent defect is the lack of a thin line 
around the ball of yolk. 
Observation shows that this defect leads to death and blackening 
within a few hours,
\qee{3} Once the embryo assumes the shape of a fish larva the
development becomes easier to follow.
The heart begins to beat some 25 hours post fertilization (hpf)
and the flow of blood becomes gradually more visible
particularly in the yolk sac and in the tail. 
In this phase 
the dead embryos can be clearly identified and counted.
\qpar

In other words, it is for the age interval between 3hpf and 20hpf
that one has to find appropriate criteria defining death.
\qpar

In order to identify dead embryos prior to (and independently from)
their blackening
one can use the following two criteria.
\qbu Further development is stopped.
\qbu The embryo does not move. 
\qpar

The ``no development'' criterion can only be used when the development leads
to changes in shape that can be clearly identified. In this respect one
can make the following remarks%
\qfoot{In order to identify a latency in development one may try to take
successive pictures of each egg in its separate well. We tried this procedure
on a small sample of 40 eggs. It turns out that
unless it can be done
automatically, this operation is quite time consuming. Moreover the pictures
are not easy to use because when the embryo has moved and shows 
a different side
successive
pictures of the same embryo do not compare easily.}%
.
\qbu As already said,
in the evolution leading to 2,4,8 cells the changes in shape and in number
can be clearly identified. However, in all our observations we have never seen
an egg arrested in the shape of 2, 4 or 8 cells. 
At first sight this may suggest that in this age group
the death rate is very low. It is certainly low but 
one must take into account that
this initial cleavage phase lasts less than two hours, 
thus the probability
of seeing a death is naturally much lower than for a phase 
which lasts several hours.
\qbu Between 2 and 8 hpf the 
embryo has the shape of a thin border line around the
yolk; this does hardly allow a clear identification of a stop in development.
\qbu After 10 hpf the embryo takes the shape of a larva rolled up
around the ball
of yolk with a head and tail clearly visible; then, in the course of time,
the tail becomes more and more detached from the yolk. This feature can be used 
to identify an arrested development.
\qpar
Table 1 summarizes the successive steps of the measurement procedure.
%
\begin{table}[htb]

\centerline{\bf Table 1:  Operations for measuring the number of deaths}
\vskip 2mm

\vskip 5mm
\hrule
\vskip 0.8mm
\hrule
\vskip 1mm
\color{black} 
\small

$$ \matrix{
 & \hbox{Time}  & \hbox{Device} \hfill & \hbox{Operation} \hfill & \hbox{Purpose}\hfill \cr
\qtb
 & \hbox{(hpf)} &     &   & \cr
\noalign{\hrule}
\qth
 1 & (0,0.2) & \hbox{Several breeding boxes} \hfill & \hbox{Production of eggs}\hfill&
 \cr
  &  & \hbox{} \hfill & \hbox{(some 100-150 per batch)}  \hfill & \cr
2 & (0.2,1.5) & \hbox{Stereomic. (x20)} \hfill & \hbox{Examination of all eggs} \hfill & 
\hbox{Separation of eggs with 1 hump} \hfill \cr 
 &  & \hbox{} & \hbox{} \hfill & 
\hbox{from those with 2 (or more)} \hfill \cr 
3 & (1.5,3) & \hbox{96 well plate} \hfill & \hbox{Repartition: 1 egg per well}\hfill &
\hbox{To follow the development of}\hfill \cr
 &  & \hbox{} \hfill & \hbox{}\hfill &
\hbox{the eggs individually}\hfill \cr
4 & (3,12h) & \hbox{Light pad +} \hfill & \hbox{Observation of all eggs} \hfill &
\hbox{Identification of black eggs}\hfill \cr
 & \hbox{} & \hbox{x10 magnifying glass} \hfill & \hbox{} \hfill &
\hbox{}\hfill\cr
5 & \hbox{{After 12h}} & \hbox{Stereomic. (x50)} \hfill &
\hbox{Identification of late comers} \hfill &
\hbox{Did development stop?}\hfill\cr
6 & \hbox{After 24h} & \hbox{Stereomic. (x50)} \hfill & \hbox{Examination of all eggs} \hfill &
\hbox{Identification of the embryos}\hfill\cr
 & \hbox{} & \hbox{} \hfill & \hbox{} \hfill &
\hbox{whose heart is not beating}\hfill\cr
\qtb
7 & \hbox{After 60h} & \hbox{Light pad} \hfill & \hbox{Identification of hatched eggs} \hfill &
\hbox{}\hfill\cr
\noalign{\hrule}
} 
$$
\vskip 0.5mm
Notes: ``hpf'' means ``hours post fertilization. 
Two modes of observation were alternated.
The light pad surveys were fast (for a 96-well plate, it took 5mn instead of 30mn for 
a microscopic observation of all the eggs),
and therefore they could be repeated often.
However, microscopic examination was necessary to identify
the eggs which were on hold and to see whether or not their heart was still beating.
The identification of laggards gives a first signal that something is wrong. It can
lead to death or to recovery.
{\it  \qL }
\vskip 2mm
\hrule
\vskip 0.7mm
\hrule
\end{table}
%

\qI{Experimental results for embryonic mortality rates} 

The goal of our experiment was to measure the mortality rate of fertilized 
embryos as a function
of the post fertilization age of the embryos. The curves which summarize the
results are presented in Fig.6a and 6b. 

%
\begin{figure}[htb]
\centerline{\psfig{width=10cm,figure=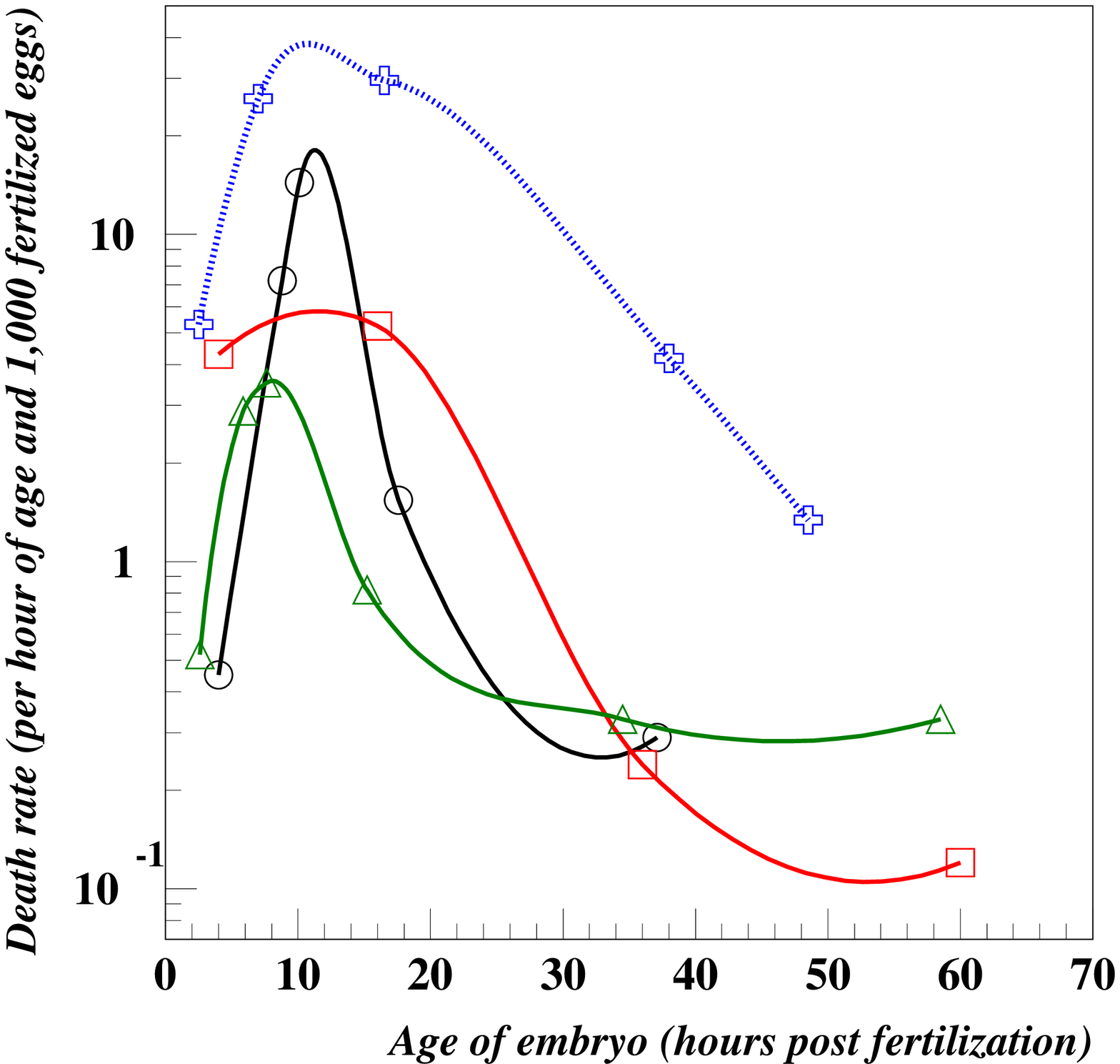}}
\qleg{Fig.\qhu 6a\qhv Embryonic mortality rates for fertilized zebrafish embryos.}
{Each curve is for a separate series of observations.
The numbers of fertilized embryos were as follows. (1) curve with circles: $ 277 $, 
(2) squares: $ 345 $, (3) triangles: $ 379 $, (4) crosses: $ 672 $ 
(total number=$ 1,673 $).
Due probably to the young age of the parents (meaning that they have
not yet
been used for reproduction, but in fact the real reason of the
fragility does not matter)
,
the 4th sample involved a high percentage of ``fragile'' eggs which led to
a huge screening process. It is to draw attention to this difference that this
curve was drawn with a dotted line. However, it is interesting to observe that
despite the difference in mortality levels, the shape of the curve is the same.
Note that the vertical scale is logarithmic. 
When necessary age intervals were merged to avoid intervals
with zero deaths.
All experiments were performed
at 23 degree Celsius. This temperature was quite stable because it was
the room temperature in a laboratory with air conditioning and
deprived of windows.
The graph shows death rates computed with the same definition
as for infant rates, i.e. number of deaths in a given age interval
divided by number of initial embryos.}
{Source: The experiments were performed during the summer of 2019 partly
at the ``Department of Life Science'' of ``Peking University''
for the initial separation of fertilized eggs from non fertilized eggs and partly
at the ``Institute of Physics'' of the ``Chinese Academy of Science'' for subsequent
observations of the embryos.}
\end{figure}
%
\begin{figure}[htb]
\centerline{\psfig{width=10cm,figure=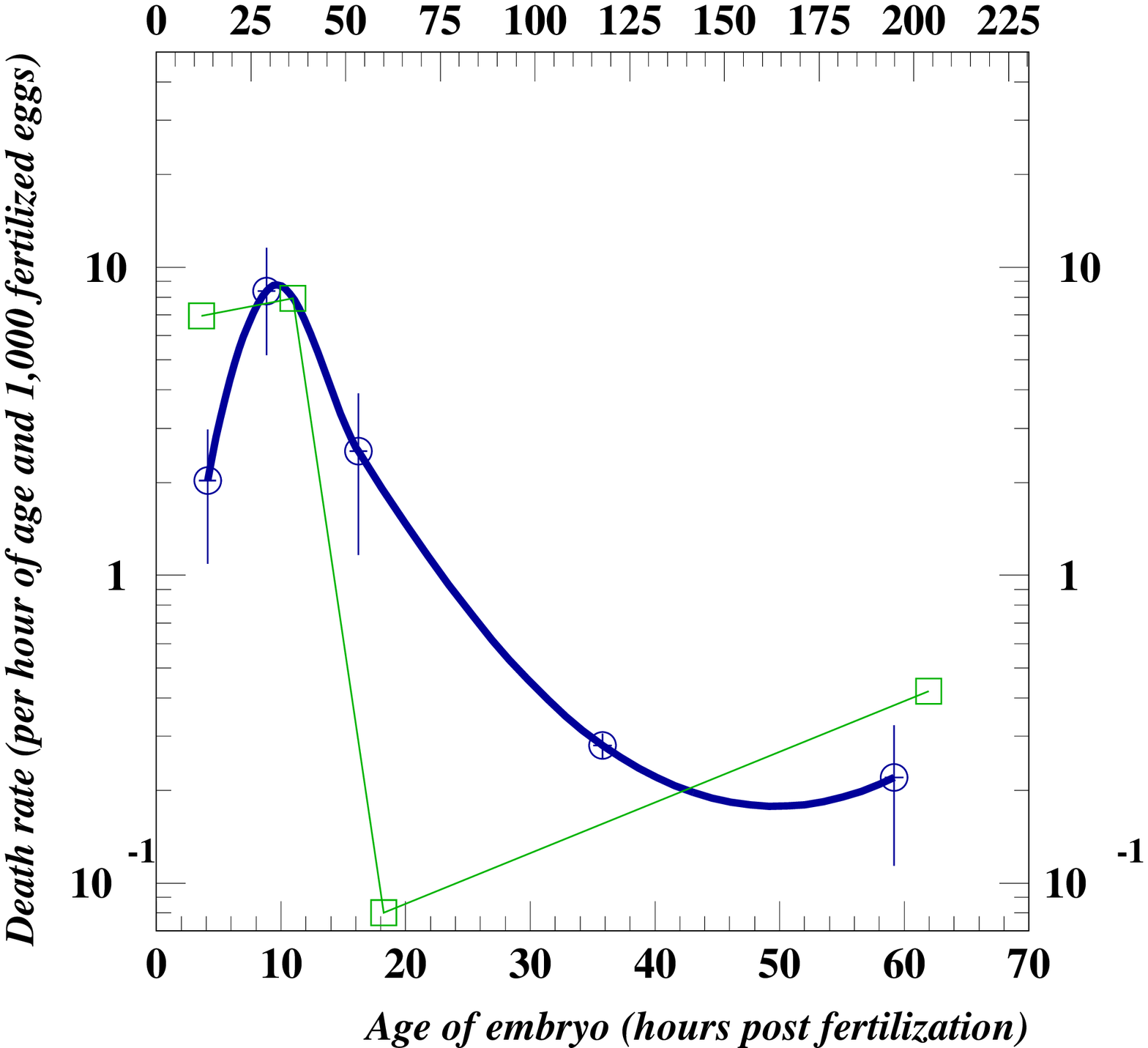}}
\qleg{Fig.\qhu 6b\qhv Average embryonic mortality rates for zebrafish
and European perch.}
{The curve with the circles is the average of the three curves 1,2,3 of 
Fig.6a and it corresponds to the lower horizontal scale
(the whole embryonic phase lasts 3-4 days). The error bars are $ \pm \sigma $ 
which defines the confidence intervals with a probability level
of 0.66. This means that if repeated 100 times under identical
conditions the experiment would give results falling within
the error bars in 66 cases.
The curve with the squares is for the European perch ({\it Perca fluviatilis})
and corresponds to the upper horizontal scale. 
Note that drawing a smooth line through this sharp 
peak was hardly possible which is why we drew straight connecting segments.}
{Sources: Zebrafish: same as for Fig.4a; {\it Perca fluviatilis}: Alix (2016).}
\end{figure}
%

Three comments are in order.
\qee{1} In Fig.6a
despite the differences in mortality levels the shapes of the
curves are similar. This stability is reassuring and gives credibility
to the early embryonic mortality surge.
This observation comes in confirmation of previous observations made 
(i) for chicken (Pe\~nuela et al. 2018, p.6505) 
(ii) for turkey (Fairchild et al.  2002, p.262)
(iii) for European perch (Alix 2016, p.161). 
Based on her research at the ``Institute of Marine Research'' 
in Bergen, Dr. Maud Alix found that for cods
there is also clear evidence for an early embryonic mortality peak
(personal communication, results to be published shortly).\qL
The results published for zebrafish by Yui Uchida et al. (2018) go in the same direction
but the effect is less spectacular due to a small sample of embryos ($ n=72 $).
\qee{2} One may wonder what is the main factor causing variability.
The comparison of curves 1,2,3 on the one hand and curve 4 on the other hand
strongly suggests that the characteristics of the parents are essential.
On average in our experiments some 4 male-female pairs 
contributed to the production of the eggs%
\qfoot{The number of breeding boxes was usually higher, say 6 or 8, but
breeding did not occur in all boxes.}%
.
Such a low number is clearly not sufficient to achieve 
homogenization
by mixing. In fact, what made sample 4 special is that it had 6 young pairs
and only two ``normal'' pairs of which only one gave eggs. Thus, there was
almost no  homogenization.
Employing more breeding boxes will improve homogenization and will likely
eliminate the most spurious cases. \qL
Naturally, instead of achieving  homogenization by mixing one can also try to
obtain it by selection, for instance by using clones. However, like real
twins, clones are affected by their life path and may be identical
only to some extent (more details about differences in real twins are
given in the
appendix of Bois et al. 2019b).
\qee{3} How does the order of magnitude of the overall embryonic mortality compare
with the mortality of the larvae after hatching? The results presented in
Bois et al. (2019, Fig.4a) for a period of 50 days post hatching show an average of
about 0.4 death per hour and 1,000 hatched larva. This is approximately the order
of magnitude that we see on the right-hand side of the curves in Fig. 6a,b. 
Moreover, the fact that the amplitude of the curves is the same for the zebrafish
and for the European perch suggests that 
embryonic death rates are fairly
robust,
remember that the perch is a much bigger fish than the zebrafish.

\qI{Conclusion}

The measurements presented in this study show that there is a huge
mortality peak 
at the beginning of the embryonic phase. 
There can be two sources of mortality. (i) The initial
elements (i.e. oocyte and sperm cell) may be defective.
(ii) Subsequent defects may arise either in the process
of fertilization or in the early stages of the 
development (of which we have given two examples).

\qA{From assisted to autonomous embryos}

It is natural
to interpret the death surge
as a screening process which will ensure successful subsequent
development of the survivors, but why should it take place
at the beginning of the embryogenesis? 

Taking again inspiration
from the birth process in which it is the 
environment change which triggers the screening process,
we can observe that here too there is an environment change
when the embryo becomes autonomous and does no longer rely
on the support of the female organism. For instance,
a partially defective function may have been acceptable as long as 
it could be propped up through interaction with the female organism.
Once released in water the fertilized embryos must rely on
their own resources. The successful completion
of the second hump seems to signal that the embryo has
passed this first screening. 
Fig.7 extends the graphs of Fig.1 to the pre-embryonic
phase because a knowledge of this phase should give
a better understanding of the EEM.

%
\begin{figure}[htb]
\centerline{\psfig{width=12cm,figure=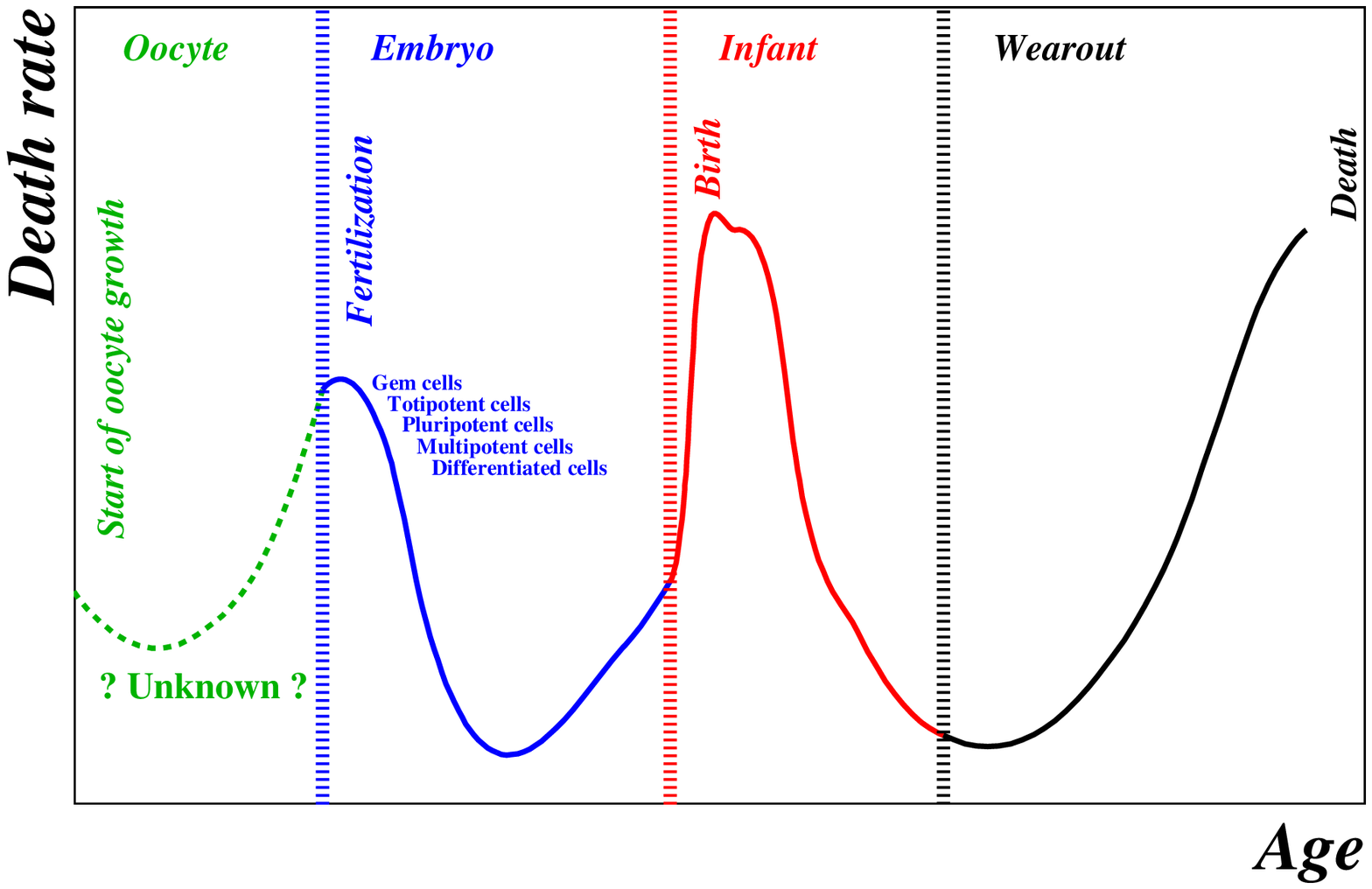}}
\qleg{Fig.\qhu 7\qhv Evolution of the death rate in the four phases of 
development.}
{This figure illustrates a conjecture which generalizes Fig.1 in two ways. 
Firstly, it 
introduces an initial phase corresponding to the ``manufacturing'' of
the spermatozoa and oocytes, respectively containing the male and
female germ cells.
Secondly, this process in four phases is presumed to apply to a broad spectrum
of species from fish to mammals and birds. However, 
it excludes organisms
whose development involves metamorphosis through successive stages (e.g.
larva, puppa, imago) which follow a more complicated process.} 
{}
\end{figure}
%

\qA{Embryonic mortality as a probe for exploring embryogenesis}

We have already mentioned the EEM measurements done by a Japanese team
of the University of Tokyo (Uchida et al. 2018) but we did not yet explain
why this is truly a pioneering investigation. It can serve as a
model for future studies in (at least) two respects.
\qbu The results for zebrafish are paralleled by similar results for chicken
and African frogs. Although essential to ensure a broad range of validity of
the results, such comparative studies are rare. The study by
Uchida et al. 
may convince biologists that the benefits justify the additional work
required by comparative investigations.
\qbu Although the title of the paper refers to a specific problem in
evolutionary embryology, its results go beyond that. 
Indeed by submitting embryos to temporary shocks%
\qfoot{Several kinds of shocks were used: temperature changes, ultra violet 
light, chemical
inhibitors, but in the subsequent discussion we focus particularly on
heat shocks.}
it was found that development is
more ``fragile'' in early than in late stages of embryogenesis,
in the sense that further development is more affected by early
than by late shocks. 
\qpar

Applying temporary or permanent temperature changes means that, 
instead of altering the ``design'' (i.e. the DNA)
one changes the conditions of ``production'' (i.e. the output of
biochemical reactions).
Such an analysis 
can be used to explore the underlying biochemical processes which 
are the components of embryonic development. Note that
this (shock $ \rightarrow $ response) methodology 
is also used extensively in 
genetics. By
observing phenotype responses to controled alterations of the DNA
one can explore the DNA and RNA sequences. 
Because such sequences can be altered
in many possible ways, these studies give an abundant harvest of data.
On the other hand, due to their wide
diversity, such results may be difficult to
integrate into a common framework.
In contrast, via  Arrhenius's law%
\qfoot{It can be written: 
$$ k=A\exp\left(-E_a/RT\right) $$
$$ k =\hbox{reaction constant},\
T=\hbox{Kelvin temperature},\ R=\hbox{constant of ideal gases},\ 
E_a=\hbox{activation energy} $$
The formula shows that the reaction constant (which controls the speed of the reaction)
increases with the temperature. At biomolecular level,
the transformations undergone by the embryo consist in biochemical 
reactions which is why biological processes
also follow Arrhenius's law.}%
,
temperature is a common determining factor of
biochemical reactions, and one can therefore
expect that temperature effects
follow fairly {\it basic} laws. 
\qpar

Finally the following objective can be mentioned
for upcoming experiments.

\qA{Mathematical characterization of the shape of the embryonic spike}

Does embryonic mortality follow an exponential or a power law decay?
A reliable answer would require more accurate death rate measurements.
As the birth peak decay is known to be
hyperbolic, a similar shape for the EEM fall
would give further confidence in a parallel between the two phenomena.

\appendix

\qI{Appendix A. Some practical issues of the experiment}

The purpose of this appendix is to explain some 
special aspects of the experiment
in the hope that it may be helpful for
researchers who wish to repeat this experiment with higher accuracy 
i.e. less background noise.

\qA{Organization of the experiment and bioethical rules}

The experiment was performed in China at the Institute of
Physics of the Chinese Academy of Science in Beijing.
The eggs were produced by standard breeding techniques
at the Life Science Department of Peking University.
Independently, a number of pictures were taken at
the ``Biology Institute of Paris-Seine'', France.

As 
zebrafish are vertebrates, according to European regulation
experimental protocols should get prior ethical approval.
However, this rule
applies only to adult fish and to larva which are
more than three days old post hatching. 
For experiments involving only embryos there are
no specific rules, hence no need for prior approval.

\qA{The oxygen supply question}

As explained in Table 1, after separation from the unfertilized,
the fertilized eggs were distributed in a 96-well plate. 
Each of these wells has a diameter of 6mm and a depth of 10mm which
gives a ratio 
$ r=\hbox{(surface of interface)/depth=2.8mm}^{-1} $.
It is this ratio which rules the oxygenation process
of the water (Chen et al. 2019).
\qpar

With respect to oxygen supply there are two distinct phases:
(i) Before the heart is working, oxygen is supplied and distributed
to the embryo by diffusion. (ii) Once the heart works, the oxygen is
still supplied by diffusion%
\qfoot{The gills start to work only some 20 days after hatching.}
but it can be distributed more effectively 
through the hemoglobin of the blood flow.
\qpar

Intuitively, it seems clear that the consumption of oxygen is correlated
with the activity of the embryo which suggests that it should become
higher as the embryogenesis progresses.  This is indeed confirmed by 
observation (Green 2004). 
While embryogenesis progresses three effects take place.
\qbu As already said the required volume of oxygen increases
\qbu Tail movements which appear on day 2 produce small displacements of the eggs and 
therefore convection through which 
the higher concentration of oxygen near the water surface in each well  will spread 
to the bottom of the well where the egg is located. 
\qbu After two days
there is a proliferation of bacteria (clearly visible in the vicinity
of the eggs) which consume oxygen.
\qpar

As it is difficult to draw a clear conclusion from
these conflicting effects
we set up a test. Instead of the 96-well plate 
we used a plate with 384 wells.
These wells have a 3mmx3mm square section and the same height of 
10mm as the 96-well plate.
This gives $ r=0.9\hbox{mm}^{-1} $, i.e. three times
smaller than for the 96-well plate. Even in such narrow wells
the embryos did not die but had a small hatching rate of only 30\%.
It can also be observed that even in the 96-plate, 
after hatching the larva
remain mostly motionless. 
This may be due to insufficient oxygen but it may also be due to
insufficient space. It is true that the larva would have 
enough space for 
swimming around the well but at this stage their natural 
movements consist in straight forward leaps for which
there is not enough space.
\qpar

Two precautions will improve the oxygenation of the water in the wells.
(1) the plate should be left open for a cover would reduce
oxygen concentration particularly in the central part of the plate.
(ii) The wells should
be filled (and refilled) with water only to one half (or one third)
of their height
so as to increase the $ r $ ratio.

\qA{The temperature question}

For zebrafish a temperature of 28 degree Celsius is generally considered as optimum.
As a consequence of the Law of Arrhenius,
the speed of the process of embryogenesis increases with temperature (at least
in the interval 20-34 degrees).
However, most of our experiments were done at a constant room temperature of 
23 degrees. Why?
%
\qbu When kept in an incubator at 28 degrees, the eggs will experience
heat shocks every time they are taken out of the incubator for observation
(and similarly when put back into it).
We do not know how such recurrent temperature shifts will affect them.
\qbu Beside the velocity effect, a
temperature under (or over) 28 degrees may also affect the mortality and
this effect may be different across age groups.
The way different age groups
respond may tell us something about the processes which take place.

\qA{Special episodes}

We mention here two special episodes. We think that they should 
be seen as over-extended manifestations of effects which occur also in
normal cases although
in less spectacular form. This is obvious for the second case and it
may also be true for 
the first.
\qpar

{\bf Latency effect}\quad In three cases we have observed a latency
time in the development process%
\qfoot{One can recall that in human embryonic development, there can be
a waiting time effect in which implantation of the embryo in the uterus 
can be delayed by several days during which the embryo remains in a
quiescent stage.
In natural conditions a distinction is made between diapause which is in response
to adverse environmental conditions and quiescence which is rather
triggered by internal factors.}
in the sense that 
during two days there was no development nor any movement of the
embryos. Then, after
this delay, in two cases the development resumed normally whereas in
the third case
the heart stopped and shortly after the egg became black.
\qpar

{\bf Effect of egg quality}\quad One set of eggs was of bad quality%
\qfoot{Due probably to the young age of the males and females.}
as was apparent by the high initial number of eggs
in bad shape (eggs of irregular shape, broken shells, eggs containing
black particles).
In such a case, if the first day acts as a filter, one expects a high 
early mortality.
This is indeed what happened. 
It is true that a high mortality for mediocre eggs is nothing surprising; 
what is more remarkable is the fact that after the first 24h their mortality rate
fell as rapidly as for normal eggs.

\count101=0  \ifnum\count101=1

\qI{Appendix B: the conundrum of Gompertz's law}

One has a clear understanding of the {\it raison d'\^etre}
of the infant mortality peak and, as explained earlier 
it is reasonable to assume a similar rationale for the
embryonic peak. On the contrary, the Gompertz's shape of 
the death rate in old age is a conundrum. 
%
\begin{figure}[htb]
\centerline{\psfig{width=8cm,figure=soap.eps}}
\qleg{Fig.\qhu B1\qhv Age-specific death rate for soap films.}
{The soap films are contained in a glass tube of a diameter of about
2cm which is closed on one side. The length of the
life-time  increases with the length of the tube due to
the influence of humidity but the 
shape of the death rate curve is conserved. 
There are two regimes: a Gompertz-like segment is followed 
by a level section which denotes a fairly constant death rate.}
{Source: The graph is based on the data of
Bois et al. (2020) (to be published).}
\end{figure}
%
In old age 
the most apparent factor is the accumulation of various
diseases: cancer, heart disease, Parkinson's, Alzheimer's
diseases and many others. 
However, why is there such an accumulation in old age%
\qfoot{This has been a matter of study in evolutionary biology. 
According to the so-called ``disposable soma theory'' reviewed by 
Kirkwood and Rose (1991) there is a point in life at 
which body maintenance does 
not compensate costs in terms of biological fitness. After this point
an increase in the mortality rate due to wear out is expected.
However, unless one can quantify notions such as
maintenance cost and body fitness this remains
a qualitative model.}%
.

While to ensure a bounded life-time
the death rate just needs to be positive
(no matter how small),
the actual death rate is not only increasing with age,
but increasing exponentially. 
Gompertz's law ensures that the human life span
cannot extend beyond 125 years (Richmond et al. 2016).
Why is that so?
What can be learned from comparative analysis?
\qpar

First of all, Fig. B1 shows that a Gompertz-like increase is
nothing exceptional since it can also be observed for 
simple systems such as soap films.\qL
Secondly, in contrast with infant mortality,
a cross-species analysis of old age mortality shows a great
diversity in shapes. Thus, comparative analysis may not
be enlightening. 
\qpar

We certainly
do not wish to claim that all characteristics of living
organisms have in some way been optimized through the
evolution process. 
For instance, in their seasonal 
migrations some bird species fly 10,000km to the other
side of the Earth (at great cost in terms of losses)
when in fact appropriate resources would
be available much closer. So, we should perhaps be ready to
accept that the 125 year boundary is nothing but an accident of
evolution without any deep significance.

\fi

\vskip 4mm

{\bf \color{blue} Acknowledgments} \quad 
We wish to express our sincere gratitude to the following persons for their
help and interest: Maud Alix of the ``Institute of Marine Research'' 
in Bergen, Norway;
Alex Bois of ``Sorbonne University''; Naoki Irie and Yui Uchida from
the ``University of Tokyo''; Xiaolong Fan, Xueri Ma and Yi Zhang from
``Beijing Normal
University''; Bo Zhang and Christopher Krueger from ``Peking University'';
Chao Jiang of the ``Institute of Physics'' of the ``Chinese Academy of Science''.


%

\vskip 4mm
{\bf References} 

\qparr
Alix (M.) 2016: Etude de la variabilit\'e de l'embryog\'en\`ese
chez la perche commune: d\'eveloppement d'approches alternatives.
[Embryogenesis of the European perch (Perca fluviatilis):
alternative approaches].
PhD Thesis presented at the University of Lorraine on 15 December
2008.

\qparr
Bell (G.), Mooers (A.O.) 1997: Size and complexity among multicellular organisms.
Biological Journal of the Linnean Society 60,345-363.

\qparr
Berrut (S.), Pouillard (V.), Richmond (P.), Roehner (B.M.) 2016:
Deciphering infant mortality.
Physica A 463,400-426.

\qparr
Bois (A.),
Garcia-Roger (E.M.),
Hong (E.),
Hutzler (S.),
Ali Irannezhad (A.),
Mannioui (A.),
Richmond (P.),
Roehner (B.M.),
Tronche (S.) 2019a: 
Infant mortality across species.
A global probe of congenital abnormalities.
Physica A 535,122308.

\qparr
Bois (A.),
Garcia-Roger (E.M.),
Hong (E.),
Hutzler (S.),
Ali Irannezhad (A.),
Mannioui (A.),
Richmond (P.),
Roehner (B.M.),
Tronche (S.) 2019b: 
Congenital anomalies from a physics perspective. The key role of
``manufacturing'' volatility.
Physica A 537,122742.

\qparr
Bois (A.),
Garcia-Roger (E.M.),
Hong (E.),
Hutzler (S.),
Ali Irannezhad (A.),
Mannioui (A.),
Richmond (P.),
Roehner (B.M.),
Tronche (S.) 2020: 
Physical models of mortality. 
Implications for defects in biological systems.
Preprint March 2020, Trinity College, Dublin.

\qparr
Chen (Q.), Di (Z.), Richmond (P.), Roehner (B.M.),  
Zhou (X.), Xu (H.) 2019: 
Basic rules governing the time scales of dissolution
of O$ _2 $ molecules in water, with a focus on biological applications.
Preprint, Beijing Normal University (November 2019).

\qparr
Cruz (M.), Garrido (N.), Herrero (J.), P\'erez-Cano (I.), Meseguer (M.)
2012: Timing of cell division in human cleavage-stage embryos is
linked with blastocyst
formation and quality.
Reproductive BioMedicine Online 25,371-381.

\qparr
Fairchild (B.D.), Christensen (V.L.), Grimes (J.L.),
Wineland (M.J.), Bagley (L.G.) 2002: Hen age relationship with
embryonic mortality and fertility in commercial turkeys.
The Journal of Applied Poultry Research 11,3,260-265.

\qparr
First (N.L.), Eyestone (W.H.) 1988: Reproductive efficiency in domestic animals.
Relative efficiency of each step in reproduction and causes of variation.
Annals of the New York Academy of Science 541,1,697-705.

\qparr
French (F.E.), Bierman (J.M.) 1962: Probabilities of fetal 
mortality. Public Health Reports 77,10,835-847.

\qparr
Glaser (O.) 1924: Temperature and forward movement of paramecium. 
The Journal of General Physiology 7,2,177-188.

\qparr
Gompertz (B.) 1825: On the nature of the function expressive of the law of 
human mortality, and on the mode of determining the value of life contingencies. 
Philosophical  Transactions of the Royal Society 115,513-585.

\qparr
Green (B.S.) 2004: Embryogenesis and oxygen consumption in benthic egg
clutches of a tropical clownfish, {\it Amphiprion melanopus} (Pomacentridae).
Comparative Biochemistry and Physiology Part A,138,33-38.

\qparr
Grove (R.D.), Hetzel (A.M.) 1968:  Vital statistics rates in the United
States, 1940–1960. 
United States Printing Office, Washington, DC.

\qparr
Hutt (F.B.) 1929: Studies in embryonic mortality in the
fowl. I. The frequency of various malpositions of the chick embryo
and their significance. Proceedings of the Royal Society
of Edinburgh 49,II,118-130. 

\qparr
Jarvis (G.E.) 2017: Early embryo mortality in natural human reproduction:
what the data say. 
FT1000Research p.1-43.

\qparr
Kirkwood (T.B.L.), Rose (M.R.) 1991: 
Evolution of senescence: late survival sacrificed for reproduction.
Philosophical Transactions: Biological Sciences, Vol. 332, No. 1262, 
``The Evolution of Reproductive Strategies'', pp. 15-24

\qparr
L\'eridon (H.) 1973:  Aspects biom\'etriques de la f\'econdit\'e humaine.
Presses Universitaires de France, Paris.\qL
[An English (somewhat expanded) translation was published in 1977 under
the title given in the following reference.]

\qparr
L\'eridon (H.) and Judith Helzner (translator) 1977: Human
fertility. The basic components. 
Chicago University Press, Chicago. Particularly: 
Ch.3: Fecundability (p.22-47) and Ch.4: Intra-uterine mortality (p.48-81).

\qparr
Pe\~nuela (A.), Hernandez (A.) 2018: 
Characterization of embryonic mortality in broilers.
Revista MVZ (Medecina, Vetenaria,Zootecnia) C\'ordoba 23,1,6500-6513.

\qparr
Post (J.), Huang (C.Y.), Hoffman (J.) 1963: The replication time and
pattern of the
liver cell in growing rat.
The Journal of Cell Biology 18,1-12.

\qparr
Richmond (P.), Roehner (B.) 2016: Predictive implications of Gompertz’s law.
PhysicaA 447,446–454.

\qparr
Richmond (P.), Roehner (B.) 2019a: A physicist's view of the
similarities
and differences between tuberculosis and cancer.
Physica A 534,1-18.

\qparr
Richmond (P.), Roehner (B.M.), Wang (Q-H) 2019b: 
The physics of large-scale food crises.
Physica A 522,274–293.

\qparr
Rideout (R.M.), Trippel (E.A.), Litvak (M.K.) 2004:
Predicting haddock embryo viability based on early cleavage patterns.
Aquaculture 230,215-228.

\qparr
Romanoff (A.) 1949: Critical periods and causes of death in
avian embryonic development.
The Auk 66,3,264-270.

\qparr
Sreenan (J.M.), Diskin (M.C.) 1986: The extent and timing of
embryonic mortality in the cow. p.1-11 of : ``Embryonic mortality in farm animals''.
edited by J.M. Sreenan and M.G. Diskin. Martinus Nijhoff.

\qparr
Tarkowski (A.J.K.)  1959: Experiments on the development of isolated blastomeres 
of mouse eggs. Nature 184,1286-1287.\qL
[This was the first experimental proof of the notion of totipotent cells]

\qparr
Uchida (Y.), Uesaka (M.), Yamamoto (T.), Takeda (H.), Irie (N.) 2018: 
Embryonic lethality is not sufficient to explain hourglass-like
conservation of vertebrate embryos.
EvoDevo 9,7,1-11.

\qparr
Wallden (M.), Fange (D.), Lundius (E.G.), Baltekin (\"O.), Elf (J.) 2016:
The synchronization of replication and division
cycles in individual {\it E. coli} cells.
Cell 166,729-739.

\qparr
Wathes (D.C.),  Diskin (M.G.) 2016: 
Reproduction, events and management: mating management: fertility
Reference Module in Food Science, 2016.

\end{document}